\documentstyle[prl,aps,preprint]{revtex}

\newcommand{\beq}{\begin{equation}}
\newcommand{\eeq}{\end{equation}}
\newcommand{\beqa}{\begin{eqnarray}}
\newcommand{\eeqa}{\end{eqnarray}}
\newcommand{\beqar}{\begin{eqnarray*}}
\newcommand{\eeqar}{\end{eqnarray*}}

\def \s {\,\,\,\,}

\begin{document}
\title{
Area scaling entropies for gravitating systems
}
\author{Jonathan Oppenheim}
\address{
Theoretical Physics Institute, University of Alberta, 412 Avadh Bhatia
Physics Lab.,\\ Alta., Canada, T6G 2J1.  Email: {\it jono@phys.ualberta.ca}
}
\date{\today}
\maketitle

\begin{abstract}
The entropy of a spherically symmetric distribution of matter in
self-equilibrium is calculated.
When gravitational effects are neglected, the entropy of the system is
proportional to its volume.  As effects due to gravitational self-interactions become
more important, the entropy acquires a correction term and is no longer
purely volume scaling.
In the limit that the boundary of the system approaches
its event horizon, the total entropy of the system is proportional
to its area.  The scaling laws of the system's thermodynamical quantities
are identical to those of a black hole, even though the system does not
possess an event horizon.
\end{abstract}


%
The entropy of ordinary matter is usually
proportional to its volume and depends crucially on
its composition.
The entropy of a black hole\cite{bek}\cite{hawk} is therefore very mysterious,
in part because it is proportional to the black hole's area $A$.
In order to explain this peculiar behavior, researchers have invoked
holography, string theory, entanglement entropy
\cite{entang}, and brick-wall models \cite{t'hooft}.

However, the volume scaling properties for the entropy of ordinary matter
depend on the fact that
interactions are short range in comparison to the size of the system, or
that the interactions are screened.  For long-range interactions such as
electromagnetism, the entropy still scales as the system's volume as long as
the net charge of the system is neutral \cite{liebandliebowitz},
but little is known about the thermodynamic properties in other cases
where long range interactions are present.

In this paper, we will examine a system where long range gravitational
interactions are present.
Although black hole entropy appears to possess very peculiar
properties, we will see that many of these features are present in
ordinary self-gravitating systems which do not possess an event horizon.
In particular, we will study spherically symmetric matter distributions
which are in self-equilibrium.  In the absence of gravity, the entropy of
the distribution is assumed to be proportional to its volume.
However, when the system's own weak gravitational field is taken into
account, the entropy acquires a correction term, spoiling its volume-scaling
properties.

In the strong field limit, the results are even
more surprising.  In the limit that the system is about to form a
black hole, its entropy is proportional to its area for a large
class of equations of state.  The scaling laws of the system's temperature
and energy are also identical to those of a black hole despite the fact
that no horizon is present.  This suggests that many of the peculiar properties
of black hole thermodynamics are the result of gravitational self-interactions,
rather than the horizon.

In this work, the system is modeled
as a dense series of thin shells of arbitrary composition.
This allows the system to be
compressed to a size close to its own event horizon \cite{fluid}.
The work
in this paper is related to reference \cite{israeletal} where
a single shell is considered.  In that case, the entropy is assumed
to always be proportional to the system's area, and when the shell
is brought close to its event horizon, the constant of proportionality
is found to be $1/4$.



%

We consider $n$
densely packed shells, where the radius of the i'th shell is given by
$r_i$, and its thermodynamical quantities, as measured by local
observers on each shell are given by
$E_i$ (mass-energy), $T_i$ (temperature), $S_i$ (entropy),
$P_i$ (tangential pressure), $N_i$ (particle number),
and $\mu_i$ (chemical potential).  The entire configuration
is assumed to be in equilibrium with itself, and therefore, it supports
itself in its own gravitational field.  If the material has a rest mass,
then this corresponds to a constant term in the chemical potential.
One could also consider an arbitrary number of particle species, however
the effect of this on our calculation is trivial.

We assume that in the absence of gravity, the system has no unscreened
long-range interactions and is therefore extensive.
In other words, the entropy, particle number and energy scale
as the size of the system when the intensive variables (temperature,
pressure and chemical potential) are held fixed.
%
In most
applications
of thermodynamics this is almost always assumed to be the case
(at least implicitly), for without it, taking the
thermodynamic limit becomes problematic.

%

We will first calculate the local thermodynamical quantities in terms of
observables at infinity.  Spherical symmetry dictates that the metric outside the i'th shell is
\beqa
ds^2 & =& {_i g}_{\mu\nu}dx^\mu dx^\nu \nonumber\\
&=&-c_i k^2_i(r)dt^2 + k^{-2}_i(r)dr^2
+ r^2 (d\theta^2+\sin^2{\theta}d\phi^2)
\eeqa
%
where throughout this paper we use units $G=c=\hbar=k_B=1$.
In order for the time coordinate $t$ to be continuous throughout the
space time, the $c_i$  need to satisfy the condition
\beq
c_i k^2_i(r_{i+1})=c_{i+1}k^2_{i+1}(r_{i+1})
\label{eq:c}
\eeq
(an overall constant is not physically relevant, and just scales the time
coordinate).
If ${_i T}_{ab}$ is the intrinsic stress-energy tensor on the i'th shell,
then its components are
related to the junction conditions of the extrinsic curvature and
intrinsic metric
\cite{mtw}
\beq
8\pi {_i T}_{ab}={_i[g_{ab}K-K_{ab}]}
\eeq
where $a=0..2$ and the notation ${_i[}\,\, ]$ indicates the jump in the bracketed quantity
evaluated at $r_i$ e.g. ${_i [}k(r)]\equiv {_i k}(r_i)-{_{i-1}k}(r_i) $.
Since the shells are static, the local energy $E_i$ of the i'th shell
is $- 4\pi r_i^2 {_i T}^t_t $, while the
tangential pressure $P_i$
is ${_i T}_\phi^\phi={_i T}_\theta^\theta$.
The jump condition thus give
\beq
E_i=-{_i[}k(r)] r_i
\eeq
\beq
8\pi P_i={_i \left[ k^\prime (r) +\frac{k(r)}{r}\right]}  \,\,\,\, .
\eeq
If we take the material to have zero net charge (as indeed we must, for
the system to be locally extensive), and have no angular momentum, then
the $k_i(r)$ are of the Schwarzschild form
\beq
k_i(r)=\sqrt{1-\frac{2 m_i}{r}} \label{eq:k} \,\,\,\, .
\eeq
Using (\ref{eq:k}) we can then find ${_i g}_{tt}$ by calculating
the constants given in (\ref{eq:c})
\beq
c_i=\frac{(r_{i+1}-2m_{i+1})(r_{i+2}-2m_{i+2})\cdots(r_{n}-2m_{n}) }
{(r_{i+1}-2m_{i})(r_{i+2}-2m_{i+1})\cdots (r_{n}-2m_{n-1}) }
\label{eq:ci}
% ...
\eeq
(with $c_n=1$).

The energy, and pressure of each shell are then
\beq
E_i=\sqrt{r_i}\lgroup \sqrt{r_i-2 m_{i-1}} - \sqrt{r_i-2 m_i}\rgroup
\label{eq:localenergy}
\eeq
\beq
P_i=\frac{m_{i-1}-r_i}{8\pi r_i^2 k_{i-1}(r_i)}
-\frac{m_{i}-r_i}{8\pi r_i^2 k_{i}(r_i)} \,\,\,\, .
\eeq

The latter equation can be thought of as the condition for mechanical
equilibrium, and ensures that the pressure balances the effect of gravity
so that the configuration is at least momentarily static.
The conditions for thermodynamic and chemical equilibrium
are the Tolman relation \cite{tolman}
\beq
T_i=\frac{T_\infty}{\sqrt{-{_i g}_{tt}(r_i)}}\,\,\,\, .
\label{eq:tolman}
\eeq
and the equality of the red-shifted chemical potentials
\beq
\mu_i \sqrt{{_ig}_{tt}} = \mu_j \sqrt{{_j g}_{tt}} \,\,\,\, .
\eeq
The latter ensures that there will be no net particle flow (one
can imagine that particles can hop from shell to shell via wires,
or simply by jumping).  The Tolman relation fixes the temperature
of the shells with respect to the temperature at infinity $T_\infty$
which we assume to be finite (to be
rigorous, one could imagine a reservoir at infinity which is attached
via a heat-conducting wire to all the shells).



These thermodynamical variables can be related to each other
by the Gibbs-Duhem relation \cite{gd}
%
%
\beq
E_i=T_i S_i-A_i P_i +\mu_i N_i \label{eq:gd}
\eeq
where $A_i$ is the area of each shell.  This equation holds
for systems which are locally extensive.  To derive this, we hold
the intensive variables ($P_i$, $T_i$, $\mu_i$) fixed, and note that
if we scale the size of a shell by $\lambda$, the entropy and energy
will scale by the same amount.  I.e.
\beq
E_i(\lambda S_i,\lambda A_i)= \lambda E_i(S_i,A_i) \,\,\,\, .
\eeq
We can then differentiate the above expression by $\lambda$
to obtain the Gibbs-Dehum relation (which is an Euler relation
of homogeneity one).  In the presence of gravity, the equivalence
principle ensures that the Gibbs-Duhem relation
is valid as it applies to
thermodynamical variables as measured in the proper rest frame of the shell.
These variables are scalars, and are therefore frame-independent.

Substituting the expressions for the local variables into (\ref{eq:gd})
one can write the entropy of each shell as
%
\beq
S_i=
\frac{1}{2T_\infty}
\left[(3m_i-r_i)-(3m_{i-1}-r_i)\frac{k_i(r_i)}{k_{i-1}(r_i)}
\right] \sqrt{c}_i
%
%
%
- \frac{\mu_n N_i}{T_\infty}\sqrt{1-\frac{2m_n}{r_n}}
 \,\,\,\,\,\,\,\,
\label{eq:ithentropy}
\eeq
%
%
%
and the total entropy of the system is just the sum
\beq
S=\sum_i S_i \label{eq:entropysum}  \,\,\,\, .
\eeq

When gravitational effects are negligible, i.e. the configuration of shells
is large in comparison to its Schwarzschild radius,
$S_i$ can be expanded to first order in $m_i/r_i$.  It is instructive
to look at some of the terms in equation (\ref{eq:gd}) individually.  The local
energy can be expanded to give
\beq
E_i \simeq m_i-m_{i-1} + \frac{m_i^2-m_{i-1}^2}{2 r_i}.
\label{newtmass}
\eeq
$m_i-m_{i-1}$ can be interpreted as the additional energy
(as measured at infinity) required to add the i'th shell to the
configuration.  Since to first order, $m_i$ is just the sum of all
$E_j$, $j\leq i$,
we see that we have recovered the usual Newtonian expression
\beq
m_i-m_{i-1}
=E_i + \phi_i
\eeq
where $\phi_i$ is the Newtonian gravitational potential
\beq
\phi_i=
-E_i\frac{E_1+E_2+...E_{i-1}}{r_i} - \frac{E_i^2}{2r_i} \,\,\,\, .
\eeq

To first order in $m_i/r_i$, the pressure term also gives the usual
Newtonian expression
\beq
P_i = -\frac{\phi_i}{2 A_i}
\eeq
The chemical potential term
\beq
\mu_n\sqrt{1-\frac{2m_n}{r_n}}
\simeq \mu_n(1-\frac{m_n}{r_n})
\eeq
contains a correction which is equivalent to the Newtonian
chemical potential due to an external gravitational field,
although in this case, the gravitational field couples not only to
the mass of each particle,
but to the entire chemical potential $\mu_n$.


One can also expand the expression for the temperature, and after
substituting these terms in (\ref{eq:ithentropy}) and taking the sum over
$i$, we arrive at the total entropy.  To zeroth order
\beq
S
\simeq
T_\infty^{-1}(
m_n-\mu_n N) \,\,\,\,
\eeq
where $N$ is the total number of particles.
We see that when we ignore the first order effects due to gravity,
the entropy, when expressed in terms
of observables at infinity, is given as
an Euler relation of homogeneity one.  This demonstrates that the system
is purely extensive.  When the intensive variables ($T_\infty$ and $\mu_n$) are held fixed, the
entropy is proportional to the total energy and total number of particles.
If all the shells have the same energy density and number density, the entropy would scale as the
volume
of the system, where the volume is understood to be the sum of all the $A_i$.

However, if we include the first order terms, we find
\beq
S
\simeq
T_\infty^{-1}(
m_n-\mu_n N + \frac{\phi}{2}  -
\mu_n N \frac{m_n}{r_n})
\label{eq:firstorder}
\eeq
where $\phi$ is the total gravitational energy required to
construct the system
\beq
\phi=\sum_i \phi_i             \s .
\eeq
The entropy includes correction terms, due to the gravitational self-interactions,
and is no longer purely extensive.

We now go to the extreme limit when the outer shell approaches
its own
Schwarzschild radius ($r_n\rightarrow 2 m_n$).  For the purposes of this paper, we will assume that
our system can approach its Schwarzschild radius, although more physically reasonable
models may not be able to generate such intense pressure or exist at such high temperatures.
Inspecting equation (\ref{eq:ci}) we see that
all the $c_i$ except $c_n$ are zero
because of the
$r_N-2m_N$ term. As a result, from equation (\ref{eq:ithentropy}) we see that all the $S_i$ except $S_n$ are zero,
although for now, we will retain the chemical potential terms.
Substituting this result in (\ref{eq:entropysum})
we can easily sum over $i$ and calculate
the total entropy
\beq
S=\frac{m_n}{2T_\infty} - \frac{\mu_n N}{T_\infty}\sqrt{1-\frac{2m_n}{r_n}}
\label{eq:swithchem}              \s .
\eeq

Pretorius, Vollick and Israel consider specific and general equations
of state where the chemical potential term in equation (\ref{eq:swithchem}) vanishes entirely \cite{israeletal},
However, we can also demand the less restrictive condition that the
chemical potential not be overly divergent i.e.
\beq
\lim_{r_n\rightarrow 2m_n} \mu_n N\sqrt{1-\frac{2m_n}{r_n}}
\ll m_n
\label{eq:chempot} \s .
\eeq
As a result, the
only term which contributes to equation (\ref{eq:entropysum}) is the
entropy of the final shell $S_n$:
\beq
S=\frac{m_n}{2T_\infty} \label{eq:totalanyt} \s .
\eeq
Since the total
entropy is equal to the entropy of the final shell, we see that the entropy is proportional to the area of the system.
That this is the case, can also be seen by noting that equation (\ref{eq:totalanyt})
is an Euler relation of homogeneity two.  This is
the precise relation which governs the entropy
and temperature of black holes.  Applying the first law
$dm_n=T_\infty dS$ to (\ref{eq:totalanyt}), one finds that
(in terms of a constant $\gamma$)
\beqa
S &=&\gamma m_n^2 \nonumber\\
&=& \frac{\gamma}{16 \pi} A_n \label{eq:SofA}
\eeqa
and
\beq
T_\infty = (2\gamma m_n)^{-1} \label{eq:tofm}
\eeq
where as before, $m_n$
is the total ADM mass as measured at infinity, and $A_n$ is the area
of the final shell (and therefore, of the entire system).

These relations, up to the constant $\gamma$, are identical to the ones which govern Schwarzschild black holes
in $3+1$ dimensions.
%
Here however, the entropy is not a property of a horizon,
but is the logarithm of the number of micro-states of
our system.
The temperature, which is usually considered to
be independent of the size of the system, is now inversely proportional to the
mass of the system.  This result is even more remarkable, since
it has not been derived by considering quantum fields on the space-time.
These later two equations are the central result of this paper.

It is also interesting to note that if, at the instant before the system forms a black
hole, it has the Hartle-Hawking temperature
\beq
T_{HH}=2m_n/A_n \label{eq:hh}
\eeq
then equations (\ref{eq:tofm}) and (\ref{eq:SofA})
imply that its entropy before black hole formation is
\beq
S(T_{HH},m_n)=A_n/4\s .
\eeq
This is identical to the entropy of a black hole of equivalent mass.

Indeed, Pretorius, Vollick, and Israel argue that
equation (\ref{eq:hh}) is the appropriate temperature for a shell as
$r_n \rightarrow 2 m_n$.  They consider quantum fields propagating
on the space-time, and demand that that the shell be in equilibrium
with the local acceleration temperature \cite{unruh}.

Ordinarily, the appropriate vacuum state for a spherically
symmetric distribution without a horizon is the Boulware vacuum (which has
a temperature of zero at infinity).
However, when the boundary
of our system lies close to its Schwarzschild radius, one cannot
consistently construct the Boulware vacuum, since
the stress-energy
of the quantum fields will be infinite (and negative), and will depend on
the number of fields.  These trans-plankian frequencies will result in an
enormous and incalculable back-reaction.
It is therefore unclear what the appropriate vacuum state and vacuum temperature are.
Pretorius, Vollick, and Israel demand that the quantum field be in the Hartle-Hawking
state with a temperature at infinity given by equation (\ref{eq:hh}),
however, this requires adding layers of insulation to the shells, and positive energy to
the field to
compensate for the back-reaction.

The manner in which this is done, and the resulting temperature of the
shells is not unique.
For example, for a single shell just outside its Schwarzschild horizon,
consistently constructing the Hartle-Hawking state requires adding a thin layer of insulation
to the shell.  The insulation prevents the fields inside the
shell from having a diverging temperature, which would result in an infinite back-reaction.
If the insulation is added to the outside of the shell, the temperature
of the shell can be zero, since it is shielded from the hot field just outside the shell.
If the insulation is added to the inside of the shell,
the local shell temperature will match the local temperature of the field and
will therefore diverge as $r_n \rightarrow 2 m_n$.

The resulting scaling laws we have derived apply when the system is in static equilibrium just outside
its Schwarzschild radius.  By extension, they also apply at the point
right before the system forms a black hole, if the collapse is
quasistatic (i.e. the system is at equilibrium at all points during
collapse).  Any system which forms a black hole through a collapse off of equilibrium,
will have an entropy bounded from above by equation (\ref{eq:SofA}),
since the equilibrium configuration maximizes the entropy.  Furthermore, for a system
which forms a black hole, the generalized second law \cite{bek} implies
that $ \gamma A_n/(16 \pi)$ is bounded from above by $A_n/4$, since otherwise
it would not be entropically favorable for the system to form a black hole.

We have seen that the usual volume scaling properties of a system
no longer hold when gravitational effects are taken into account.
In the limit that our system approaches its Schwarzschild radius,
its entropy scales as the area of the system.  The temperature
is also no longer an intensive variable, and scales inversely with the system's
total energy.  In other words, the system's thermodynamical quantities scale in the same
manner as those of a black hole.
These rather surprising results suggestions that at least to an
extent, some of the intriguing aspects of black hole entropy can
also be found in self-gravitating systems which do not possess
a horizon.  Once long range interactions are taken into account,
the entropy of a system can behave in unfamiliar ways, and lead
to scaling behavior which is no longer extensive.


The ability of high-energy theories such as loop quantum gravity and
string theory to make predictions about the black hole entropy has
generated much interest, and provides crucial insights into the entropy of
black holes.
The results in this work suggest that a wide class of systems will
also result in the correct entropy scaling laws for black holes.
It is hoped that by studying the entropy of these self-gravitating systems,
we can learn more about black hole entropy.  It is also hoped that
studying the thermodynamics of self-gravitating systems will lead
to a better understanding of the thermodynamics of more general non-extensive
systems.

%


\vskip .5cm
\noindent{\bf
Acknowledgments:} It is a pleasure to thank Don Page for many interesting,
and invaluable discussions.  It is also a pleasure to thank Sriramkumar Lakshmanan and
Frank Marsiglio for their helpful comments.  This research was supported in part by NSERC.


\begin{thebibliography}{9}
\bibitem{bek} J. Bekenstein, Lett. Nuovo Cim. {\bf4},1972
\bibitem{hawk} S. Hawking, Comm. Math. Phys. {\bf 43}, 199 (1975)
\bibitem{entang} M. Srednicki, Phys. Rev. Lett {\bf 71} 666 (1993)
\bibitem{t'hooft} G. t'Hooft, Nucl. Phys. {\bf B256}, 727 (1985)
\bibitem{liebandliebowitz} J. Lebowitz, E. Lieb, Phys. Rev. Lett {\bf 22}, 631 (1969)
\bibitem{fluid} The entropy of self-gravitating perfect fluid models is also not volume scaling,
J. Oppenheim, in preparation.
\bibitem{israeletal} F. Pretorius, D. Vollick, W. Israel, Phys. Rev. D. {\bf 57}, 6311, (1998)
\bibitem{mtw} C. Misner D. Sharp Phys. Rev. B. {\bf 136}, 571(1964);
W. Israel, Nuovo Cimento 44 B, 1 (1966)
\bibitem{tolman} R. Tolman, P. Ehrenfest, Phys. Rev. {\bf 36}, 1761 (1930)
\bibitem{gd}
For applications of the Gibbs-Duhem to gravitational systems,
see for example, S. Weinberg, {\it Gravitation and Cosmology},
Wiley, (1972) as well as reference [7]
\bibitem{unruh} W. Unruh, Phys. Rev. D {\bf 14}, 870 (1976)
\end{thebibliography}
\end{document}